\documentclass[times,11pt,twocolumn]{article} 
\usepackage{latex8}
\usepackage{times}

\usepackage{amsmath}
\usepackage{amssymb}
\usepackage{bm}
\usepackage{multirow}
\usepackage{graphicx}
\usepackage{textcomp}
\usepackage{xcolor}
\usepackage{url}
\newif\ifshowcomments
\showcommentstrue

\ifshowcomments
\newcommand{\mynote}[2]{\fbox{\bfseries\sffamily\scriptsize{#1}}
	{\small$\blacktriangleright$\textsf{\emph{#2}}$\blacktriangleleft$}}
\else
\newcommand{\mynote}[2]{}
\fi

\begin{document}

\title{Android Malware Family Classification\\Based on Resource Consumption over Time\\ (Extended Version)}

\author{Luca Massarelli, Leonardo Aniello, Claudio Ciccotelli,\\ Leonardo Querzoni, Daniele Ucci and Roberto Baldoni\\
Research Center of Cyber Intelligence and Information Security (CIS)\\
Department of Computer, Control, and Management Engineering ``Antonio Ruberti''\\
``La Sapienza'' University of Rome, Italy\\
Email: \{massarelli, aniello, ciccotelli, querzoni, ucci, baldoni\}@dis.uniroma1.it}

\maketitle
\thispagestyle{empty}

\begin{abstract}
The vast majority of today's mobile malware targets Android devices. 
This has pushed the research effort in Android malware analysis in the last years. 
An important task of malware analysis is the classification of malware samples into known families. 
Static malware analysis is known to fall short against techniques that change static characteristics of the malware (e.g. code obfuscation), while dynamic analysis has proven effective against such techniques. 
To the best of our knowledge, the most notable work on Android malware family classification purely based on dynamic analysis is DroidScribe.
With respect to DroidScribe, our approach is easier to reproduce.
Our methodology only employs publicly available tools, does not require any modification to the emulated environment or Android OS, and can collect data from physical devices.
The latter is a key factor, since modern mobile malware can detect the emulated environment and hide their malicious behavior.
Our approach relies on resource consumption metrics available from the \texttt{proc} file system.
Features are extracted through detrended fluctuation analysis and correlation. 
Finally, a SVM is employed to classify malware into families. 
We provide an experimental evaluation on malware samples from the Drebin dataset, where we obtain a classification accuracy of 82\%, proving that our methodology achieves an accuracy comparable to that of DroidScribe.
Furthermore, we make the software we developed publicly available, to ease the reproducibility of our results.
\end{abstract}

\Section{Introduction}

	


\medskip

The relentless growth of smartphone sales and their pervasiveness in our daily lives fostered the development of malicious software for attacking mobile devices.
%
Android OS is the most diffused platform for mobile devices~\cite{Gartner2017}, its source code is publicly available.
Moreover, the majority of Android devices still run dated versions\footnote{https://developer.android.com/about/dashboards/index.html}.
These factors combined together make Android smartphones attractive for several malware authors~\cite{tam2017evolution}.
For these reasons, the majority of mobile malware is designed to attack Android.
According to F-Secure, 79\% of mobile malware in 2013 were designed to attack this OS~\cite{F-Secure2013}.
Only in 2015, Symantec observed a $230\%$ increase of malicious Android apps with respect to the previous year and noticed the raise of mobile malware leveraging evasion techniques to avoid the detection of signature-based security products~\cite{SymantecISTR2016}.
These techniques include mobile app obfuscation and virtualized environment detection.
In particular, the adoption of obfuscation techniques has facilitated the spread of \textit{variants} of already known malware, with 40\% growth in 2015.
Obfuscation techniques modify app's packages and/or source code, preserving app original functionalities, and allow malware authors to create semantically similar applications that are syntactically different from each other (e.g. the same trojan application can be distributed under the guise of two completely different harmful software).
Combining together trivial and advanced obfuscation techniques reduces the effectiveness of anti-malware products from 95\% to 40\%~\cite{pomilia2016Obf}.


In response to the progress of malicious software, the field of \textit{malware analysis} has advanced as well along different directions~\cite{laurenza2016architecture}. In addition to \textit{malware detection}, which entails classifying software samples as either benign or malicious, other relevant research paths have emerged. One of the most interesting is \textit{malware family classification}, which consists in classifying malware into \textit{families}. Samples belonging to the same family show similar behavior, exploit the same vulnerabilities and have the same objectives, hence family classification can be used to quickly check if an app is a variant of already known malware~\footnote{It is to note that classifying malware in \textit{families} is different from categorizing them according to their \textit{type} (e.g., a dropper, or a trojan): as an example, there can be droppers of distinct families which perform the same general task (i.e., installing the actual malware somewhere in the target system) but in diverse ways (e.g., one can download the malware from the web and the other can contain itself the malicious payload and decipher it when needed), hence their behaviors would correctly result dissimilar.}.
While Symantec registered a significant growth of Android mobile malware variants, they have not identified the same trend in the number of newly discovered malware families: just a 6\% increase in 2015 over the previous year, and only 1\% in 2016~\cite{SymantecISTR2017}. Such tendency suggests an evident slowdown in the innovation of mobile malware development, which can be leveraged by analysts with the use of family classification techniques.
Indeed, \textit{automatic malware family classification} allows malware analysts to timely understand whether a malicious app is likely to be a variant of known malware (i.e., it belongs to an already known family) or if instead is a novel malware (i.e., it does not belong to any known family). This enables analysts to focus on really brand new malicious apps only, without wasting time and effort on dissecting samples similar to others already analysed in the past. \textit{Given the huge amount of new mobile malware produced every day, using family classification to filter out what apps deserve more detailed analyses becomes fundamental.}
In practice, malware analysts can perform two different types of analyses: static and dynamic.
Static approaches do not require the execution of samples under analysis and can potentially reveal all the sample's execution paths.
Nevertheless, they are not very effective against obfuscation techniques, as extensively demonstrated in~\cite{pomilia2016Obf}, and are not able to track both modifications at runtime and generated network traffic traces~\cite{tam2017evolution}.
Dynamic techniques overcome these limitations by executing samples in controlled environments. 

This paper focuses on malware family classification.
We present a methodology that relies on dynamic analysis and propose a software architecture implementing it.
The architecture automatically executes Android applications in a sandboxed environment and generates stimuli to simulate user inputs.
During each run, it collects resource consumption metrics from the \texttt{proc} file system, and processes them through detrended fluctuation analysis (DFA)~\cite{peng1994mosaic} and Pearson's correlation~\cite{pearson1895note}.
At the time of writing, the most important work on Android malware family classification, based on dynamic analysis, is DroidScribe~\cite{dash2016droidscribe}.
We compare our methodology to that of DroidScribe by carrying out an extensive experimental evaluation on the same dataset as DroidScribe.
Results show that our methodology achieves comparable accuracy, but (\textit{i}) it is easier to reproduce and (\textit{ii}) collected data can be gathered on physical devices.
Indeed, our approach only employs publicly available tools and does not require any modification to the emulated environment or Android OS.
Conversely, DroidScribe relies on CopperDroid, which is not publicly available and it is only accessible through an on-line service.
Nevertheless, it is not suitable for batch experiments, since it take in input a just one sample and submission procedure cannot be automated as it requires to pass an anti-bot challenge-response test.
At the time of writing, the service does not analyze enqueued APKs since July 2015.
In addition, differently from CopperDroid, our methodology can collect data on physical devices as well.
This is a key factor, since modern mobile malware can detect the emulated environment and hide their malicious behavior.
To the best of our knowledge our methodology is the first based only on resource consumption metrics and DFA ~\cite{tam2017evolution}.

\smallskip

\noindent The rest of the paper is structured as follows.
Section~\ref{sec:related_work} discusses the related work.
Section~\ref{sec:methodology} presents proposed methodology and architecture.
Experimental evaluation and comparison with DroidScribe are reported in Section~\ref{sec:evaluation}, while Section~\ref{sec:conclusion} concludes the paper and outlines future work.

\Section{Related Work}
\label{sec:related_work}

Android malware family classification is a well known problem in literature and differs from malware detection regarding the final objective of the analysis, which considerably affects what specific techniques are employed although in general similar approaches are used. For example, while binary classifiers are generally used for malware detection (i.e., a sample is either benign or malicious), multiclass classifiers are instead commonly employed for family classification (i.e., one class for each family), but the set of extracted features can be very similar.
Most of the works relying on Android dynamic analyses monitor APIs and system calls. To the best of our knowledge, our work is the first to address family classification  leveraging resource consumption metrics.

Karbab \textit{et al.}~\cite{karbab2016dysign} use dynamic analysis and Natural Language Processing (NLP) to detect and classify Android malware.
Mobile applications are executed into a sandbox to generate a report of their activities, later processed by NLP techniques.
By means of these techniques, the authors are able to produce a signature for identifying and classifying malicious apps.
Reina et \textit{al.}~\cite{reina2013system} implemented CopperDroid, a framework to execute Android apps and collect information about system calls.
DroidScribe~\cite{dash2016droidscribe} uses CopperDroid as building block to execute Android apps and trace performed system calls.
DroidScribe is the most notable work on Android malware family classification, purely based on dynamic analysis and machine learning, that shares most similarities with our objectives.

Similarly to our work, Shehu \textit{et al.}~\cite{shehu2016towards} use resource consumption metrics to create a fingerprint for each mobile application under analysis.
However, they address a different problem, that is detecting obfuscated malware variants. Moreover, their methodology is tested manually on a physical device over a rather small set of 7 malicious applications.
In this paper, instead, we automate the whole process of malware family classification and test our results on a larger dataset.

Other works  leverage resource consumption metrics or power consumption to detect Android malware.  
Liu \textit{et al.}~\cite{liu2009virusmeter} and Kim \textit{et al.}~\cite{kim2008detecting} look at power consumptions of mobile devices to detect malware. Anyhow, these works are based on obsolete mobile platforms and this type of analysis can be performed on device only as power consumption metrics are not meaningful when measured on emulators or simulators.
Amos \textit{et al.}~\cite{amos2013applying} implemented an automatic framework for executing and collecting resource consumption related features to feed different machine learning algorithms.
Nevertheless, they evaluated their accuracy over a test set of only 47 applications.
Canfora \textit{et al.}~\cite{canfora2016acquiring} leverage resource consumption metrics to detect Android malware. However, these works do not address family classification.

Finally, Mutti \textit{et al.}~\cite{mutti2015baredroid} developed \textit{BareDroid}, an efficient system for analysing Android Malware on device. On device analysis has the great advantage to be immune to emulator evasion techniques but is more time-consuming, since physical devices require more time to be reset to a clean state with respect to an emulator. The authors showed experimentally that BareDroid allows to reduce this time by a $4.44$ factor.


\Section{Methodology}
\label{sec:methodology}

As already discussed, this paper focuses on malware family classification. 
That is, the task is to take in input malware samples and classify each sample into its family.
We assume that the set of Android malware families $\mathcal{F}$ is fixed and known a priori.
%

Our methodology can be summarized as follows.
In a prior training stage, we collect a large set of known malware $\mathcal{M}$ in which each sample has already been labeled with the family it belongs to.
$\mathcal{M}$ must contain members of each family in $\mathcal{F}$.
In our experiments we employed the Drebin dataset~\cite{arp2014drebin}. 
For each sample $s \in \mathcal{M}$ we run $s$ into a controlled emulated environment and collect some runtime metrics over time, so as to obtain a time series for each metric.
In particular, our methodology only relies on monitoring resource consumption metrics that can be obtained from the \texttt{proc} file system.
Then, we process these metrics so as to extract the fingerprint as detailed later.
Finally, we train our classifier with the set of generated fingerprints (labeled with the corresponding family).
Whenever, a new malware sample $s$ is given in input, we run $s$ into our controlled emulated environment and we build its fingerprint. Then, we feed our classifier with the newly generated fingerprint, which, in turn, outputs the family of $s$.


\SubSection{Fingerprint Generation}\label{sec:fingerprint_generation}
The first phase of fingerprint generation for a malware sample $s$ is the execution of $s$ in our controlled emulated environment. 
During the execution we stimulate the malware with a predefined set of input events.
The input events are automatically simulated by our emulator (refer to Section~\ref{sec:implementation} for implementation details).
To make the simulation of input events consistent across different executions, we always generate the same random sequence of events.
During the execution we monitor and collect $n$ metrics from the \texttt{proc} file system over time, at a sampling interval $\tau$. In our experiments we monitored $n = 26$ metrics at a sampling frequency of 4 Hz ($\tau = 0.25$s). 
We monitor a set of system-wide and application-specific metrics, including CPU, memory and network usage. 
Thus, at the end of the execution we obtain a time series $x_i(t)$ for each metric $x_i$.
After each execution the emulator is reset to its original image, so that each new execution starts in the same runtime environment.
Then, we process the collected time series so as to extract a vector of features $\bm{f} = g(x_1(t),\dots,x_n(t)) = (f_1,\dots,f_m)$ that characterizes the malware family (where $g$ is the feature extraction algorithm). 
The feature vector $\bm{f}$ represents the fingerprint computed for the malware sample $s$.
An appropriate set of features to include in the fingerprint is determined during the training phase, as detailed in Section~\ref{sec:training}.

\SubSection{Classification and Training}\label{sec:training}
To classify the malware family we feed a support vector machine (SVM) classifier with the fingerprint computed as described in the previous section.
%
The SVM has to be previously trained on a set of sample fingerprints, labeled with the corresponding correct malware family.

The training phase serves two main purposes: (T1)~determine the set of features to be included in the fingerprint and (T2)~train the classifier.
The first task defines the feature space of the classifier. Determining an appropriate set of features is fundamental to correctly classify.

Both tasks need a training dataset. The first step of the training phase is the collection of such a dataset.
We execute each malware sample $s \in \mathcal{M}$ in our emulated environment and we collect the monitoring metrics as described in Section~\ref{sec:fingerprint_generation}.
For each malware sample, we perform $q$ runs. The larger $q$, the lower the probability that noisy data and outliers affect the classifier training (in our experiments ${q = 2}$).
At the end of this step, we get the time series ${\bm{x_s^i} = (x_1(t), \dots, x_n(t))}$ for each run $i = 1, \dots, q$ of each sample $s \in \mathcal{M}$.

For the purpose of task (T1) we first process each vector $\bm{x_s^i} = (x_1(t), \dots, x_n(t))$ through DFA. 
We have opted for DFA, in place of similar methods, because it has no parameters to tune, thus allowing for an automatic approach. 
In particular, for each time series $x_i(t)$, $i \in [1,n]$, we compute the DFA exponent $\alpha_i$.
To support our intuition that the DFA exponent may be a valuable feature for malware classification, we conducted some experiments to show the stability of such a parameter when applied to resource consumption metrics of an Android application (see section~\ref{sec:stability_dfa}).
In addition to DFA exponents, we compute the correlation matrix of $x_1(t), \dots, x_n(t)$, so that we obtain a Pearson's correlation coefficient $r_{ij}$ for each pair of distinct metrics $x_i$, $x_j$.
Thus, we end up with an initial set of $\frac{n(n+1)}{2} = 351$ features, consisting in $n = 26$ DFA exponents and $\frac{n(n-1)}{2} = 325$ correlation coefficients.

This initial set may contain features that are redundant, may not be useful to discern the malware families, or even mislead the classifier.
Thus, we perform feature selection to determine an appropriate subset of features to feed the classifier.
To this aim, we take a wrapper approach~\cite{Guyon:2003:IVF:944919.944968}, that is we exploit the classifier itself to evaluate the goodness of a particular subset of features. Given a \emph{candidate} subset of features we train the classifier with that set, and we use the training accuracy (i.e., the accuracy obtained on the validation set during training) as a measure of its goodness.
Finally, we retain the subset that allowed to achieve the best accuracy.

The candidate subsets of features are constructed as follows.
For each feature $f$ in the initial set of all features we compute the mutual information~\cite{Thomas1991} between the distribution of all samples of $f$ and the corresponding distribution of labels (i.e., malware families).
The mutual information function gives us a coefficient that quantifies how much knowing the value of the feature $f$ for a generic malware sample $s$, gives us information about the family of $s$.
Intuitively, the larger the value of the mutual information for a given feature, the most valuable it is to recognize malware families.
We consider subsets of features such that each feature has a mutual information value that is at most $Q\%$ less than the maximum value obtained.
We generate such a dataset for each $Q \in \lbrace 10, 15, 20, 25, 30, 35, 40, 45, 50 \rbrace$.
Moreover, we process each subset of features through principal component analysis (PCA) \cite{Jolliffe2002} to further reduce the feature space.

We train a SVM for each subset and finally retain the subset of features that yield the best training accuracy. 
These features are the ones that form the feature vector, i.e., the fingerprint, which completes task~(T1).
This also results in the completion of task~(T2). Indeed, the corresponding trained SVM becomes the classifier.

\SubSection{Implementation}\label{sec:implementation}


In this section we describe an implementation of our methodology that we used in our experiments.

\SubSubSection{Architecture}\label{sec:architecture}

\begin{figure}[!t]
\centering
\includegraphics[width=\linewidth]{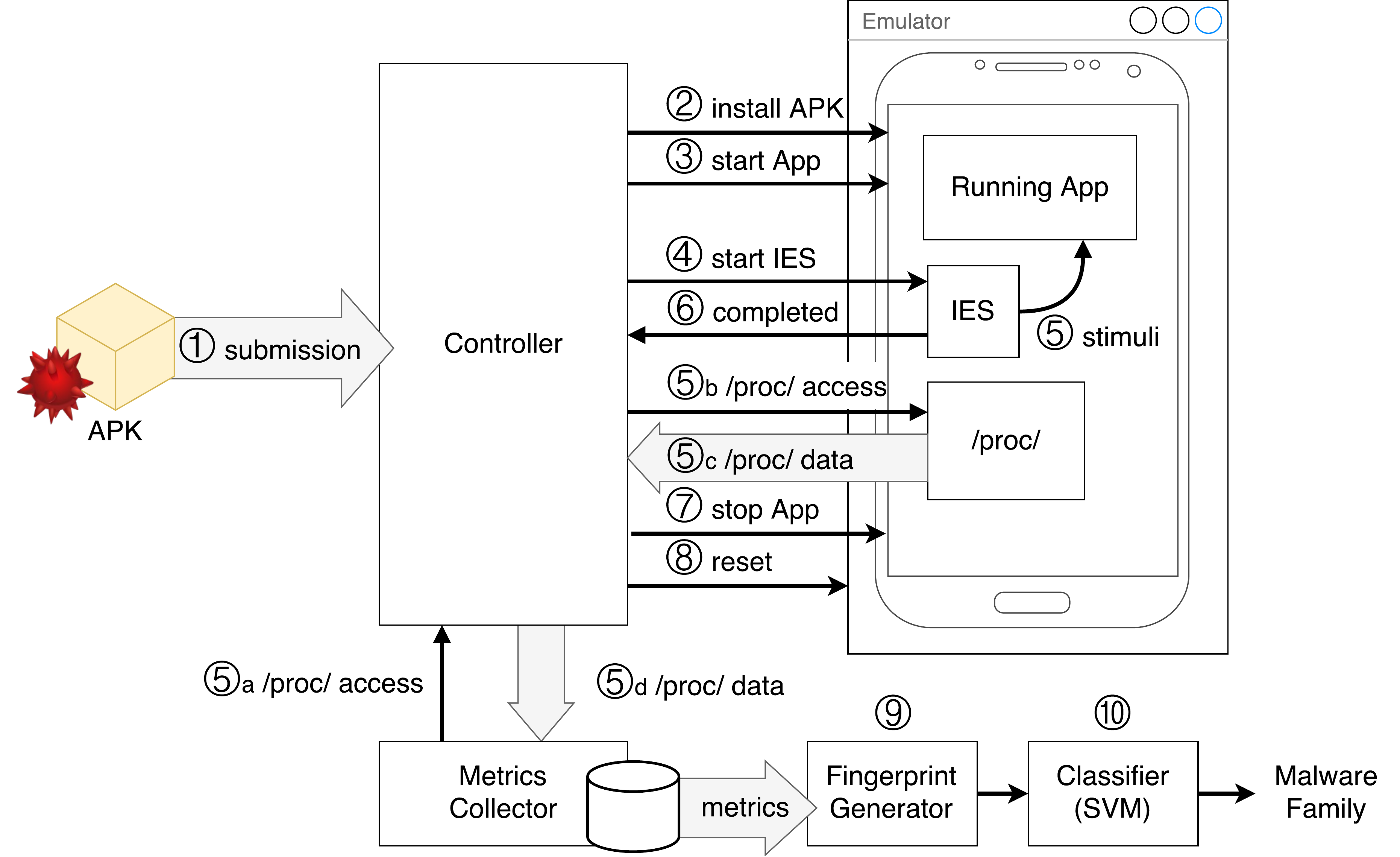}
\caption{Proposed architecture and workflow for Android malware family classification.}
\label{fig:architecture}
\end{figure}

Figure~\ref{fig:architecture} shows an architecture for our methodology and its expected workflow.
Central to the architecture are the emulator, that executes an instance of Android in which malware samples are run, and the related \textit{Controller}, that handles any interaction with the emulator.
First, when a new malware sample is submitted to the system (in form of an APK) the controller is in charge of installing the application into the emulated Android and launch it.
During the execution, the app is stimulated through a sequence of random input events by means of the \textit{Input Events Simulator} (IES) component installed in Android.
During the input stimulation phase, the \textit{Metrics Collector} component accesses the \texttt{proc} file system of the emulated instance of Android through the Controller.
Every sampling interval $\tau$, the Metrics Collector requests access to the \texttt{proc} file system through the Controller, read resource consumption metrics, and stores them in a local file.

When the IES component notifies the Controller that all input events have been simulated, the latter stops the execution of the application and reset the emulator to its initial conditions.
Finally, the \textit{Fingerprint Generator} analyzes the file containing collected metrics to extract the features and generate the fingerprint of the submitted application.
The fingerprint is, then, given in input to the SVM classifier which outputs the family of the malware sample.



\SubSubSection{Third Party Tools}
\label{sec:third_party_tools}

To implement the proposed architecture we employed several publicly available third-party tools. We briefly present each one, before discussing how to integrate them to build the architecture.

\noindent \textbf{VirtualBox\footnote{https://www.virtualbox.org/}} is an open source hypervisor.
It allows to virtualize a guest operating system on a physical machine potentially running a different OS.

\noindent \textbf{Genymotion\footnote{https://www.genymotion.com/}} is an Android emulator that leverages VirtualBox for virtual environment creation.
Genymotion provides all the features of a real mobile device, which can be controlled and monitored using VirtualBox's command-line interface.

\noindent \textbf{Android Debug Bridge\footnote{\url{https://developer.android.com/studio/command-line/adb.html}} (ADB)} is a command-line tool enabling an OS to interact with an emulated Android device and have access to all its resources.
ADB also handles installation as well as launch and termination of applications.

\noindent \textbf{UI/Application Exerciser Monkey\footnote{https://developer.android.com/studio/test/monkey.html}} is a program running on Android devices which can be invoked by the command-line through ADB.
It generates stimuli for apps running on the Android emulator by simulating a wide variety of inputs including touches, movements, clicks, system events, and activity launches.

\noindent \textbf{NOnLinear measures for Dynamical Systems\footnote{https://github.com/CSchoel/nolds} (NOLDS)} is a Python package that includes different algorithms for one dimensional time-series analysis, including DFA, sample-entropy and Hurst exponent.

\noindent \textbf{NumPy\footnote{http://www.numpy.org/}} is a Python package for scientific computing. We employ it to compute the correlation matrix.

\noindent \textbf{Scikit-learn}~\cite{scikit-learn} is a Python package that includes several machine learning algorithms for supervised and unsupervised learning, including SVMs. 

\SubSubSection{Tools Integration}

The core of our architecture, namely the emulator, is realized through Genymotion.
ADB serves as the Controller component.
It handles the installation of applications in the emulator, and can start and stop applications through the appropriate commands. 
The UI/Application Exerciser Monkey runs within Android and can simulate input events on any running application. 
Interactions with Monkey are handled by ADB.
It is possible to ask Monkey to generate a random sequence of events.
The sequence is univocally identified by a seed and thus is reproducible.
In our methodology we make Monkey always generate the same random sequence of input events.
It implements the IES component.
%
The Metrics Collector component of our architecture is implemented through Python script which periodically interacts with ADB.
%
When all input events have been simulated by Monkey, the application is stopped as well as the virtual machine.
Also, the original image of the Android virtual machine is restored through the VirtualBox command line tool.
The Fingerprint Generator is implemented as another Python script that processes the file containing the collected metrics and computes the fingerprint.
The features are computed through appropriate functions of the NOLDS and NumPy packages.
Finally, the classifier is built and trained through the Scikit-learn package.
The whole workflow described above is completely automated and handled through proper Python scripts.
All materials required to reproduce the experiments has been made available on a Github repository\footnote{https://github.com/lucamassarelli/AMFC-BRCT}.


\Section{Experimental Evaluation}
\label{sec:evaluation}

We carried out an extensive experimental evaluation to validate our approach. In this section we discuss the details of the experiments and present the results.

\SubSection{Dataset}

We performed our experiments on the Drebin dataset~\cite{arp2014drebin}. 
It is a public collection of malware samples that can be used for research purposes, that contains 5,560 malicious applications from 179 different families. 
The Drebin dataset has been widely employed in the related literature~\cite{gonzalez2014droidkin,canfora2016acquiring,dash2016droidscribe,karbab2016dysign,Suarez-Tangil:2017:DFA:3029806.3029825}. 
Since we compare our solution to DroidScribe, in our experiments we selected the same families as DroidScribe.

\begin{figure}
	\centering
	\includegraphics[width=\columnwidth]{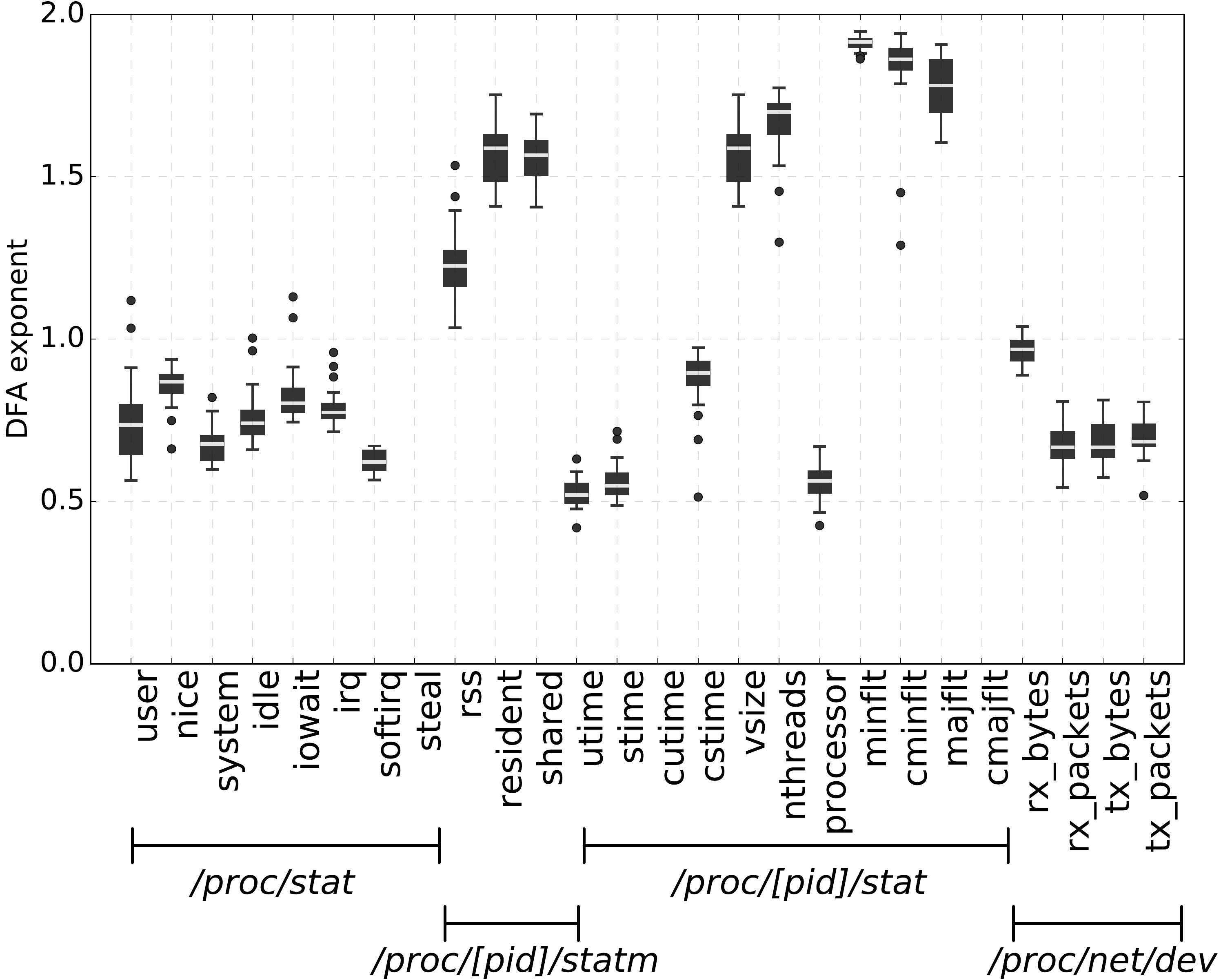}
	\caption{Box plot of the DFA exponents among 30 independent executions for a randomly sampled malicious app, for each of the 26 metrics.}
	\label{figure:stability_statistic}
	
\end{figure}

\SubSection{Experimental Setup}
We run our experiments on a physical machine with an 8 cores Intel Xeon CPU and 16 GB of RAM. All software, including the emulator (Genymotion and VirtualBox) and the Python scripts that control the workflow of the experiments run on this machine.
The emulator runs an unmodified instance of Android 4.0.
We chose this version of Android because it was released during the period in which the samples of Drebin dataset were collected. 

We implemented our methodology as described in Section~\ref{sec:implementation}.
In every execution we generate the same random sequence of simulated input events through Monkey. 
Furthermore, we configured Monkey to generate each event type with equal probability. 
In all experiments we generate $S=10,000$ input events.
This number has been determined after a preliminary study on the stability of the DFA exponent with respect to increasing numbers of generated input events.
The results of this analysis are discussed in the next section.
The stability of the Pearson's correlation coefficient between resource consumption metrics of Android apps has been already studied in~\cite{shehu2016towards}.
Every experiment is conducted in isolation as we reset the emulator after every execution.

\SubSection{Stability of the DFA exponent}\label{sec:stability_dfa}
In order to be useful, the fingerprint of an application should remain \emph{consistent} across different executions.
Despite the conditions of the experiments are kept constant across different executions, a bit of variability is inevitable, due to multiple factors that are out of our control, such as network latencies and workload fluctuations.
Therefore, we do not expect the fingerprint to remain perfectly constant across different executions. However, it is sufficient for our purposes, that the variability is bounded.

As starting point we fixed the conditions of the experiments and we assessed the variance of the DFA exponent across several executions.
We selected a random sample of malicious apps, executed each malware 30 times, collected the 26 metrics and computed the corresponding DFA exponents.
Figure~\ref{figure:stability_statistic} shows the results of these experiments for a single application. 
For each of the 26 collected metrics the figure shows the box plot of the corresponding DFA exponent.
The bottom and top of each black box represent, respectively, the first and third quartiles of the DFA exponents computed for a given metric.
The white line within each black box is the median value of the DFA exponent of the related metric.
Whiskers show the interquartile range and points are outliers.
As the figure shows, the DFA exponents are not constant across different executions, but they are characterized by a low variance.
This preliminary result supports our intuition that the DFA exponent may be an appropriate feature for the fingerprint.

\begin{figure}
\centering
\includegraphics[width=\columnwidth]{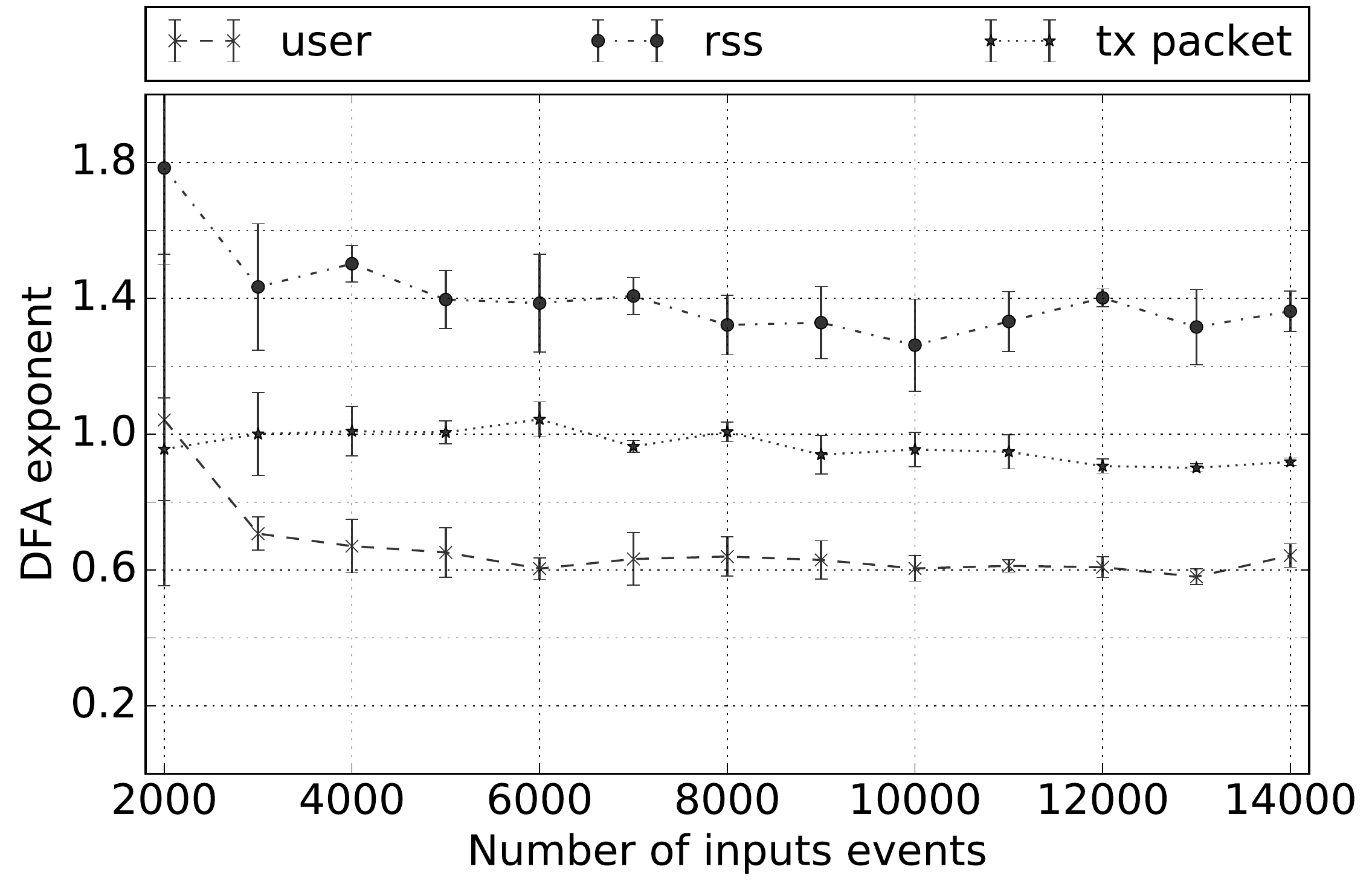}
\caption{Stability of the DFA exponent for three different metrics: \textit{User CPU}, \textit{Resident Set Size}, \textit{Transmitted Packets}.}
\label{figure:stability}
\end{figure}

Subsequently, we assessed the stability of the DFA exponent with respect to the number of simulated input events.
We performed several experiments on a random sample of malicious apps with an increasing number of stimuli $S$, ranging from $2,000$ to $14,000$ with step equal to $1,000$.
For each app and value of $S$ we performed $5$ runs, and in each run we computed the DFA exponents for all metrics.
Figure~\ref{figure:stability} shows the trend of the DFA exponent for $3$ selected metrics (user CPU usage, resident set size, number of transmitted packets) with increasing numbers of stimuli.
Each point is relative to a metric $x$ and a fixed number of stimuli $S$.
It reports the mean value of the DFA exponents computed for the metric $x$ during the 5 runs in which the number of generated stimuli was $S$.
The bars represent standard deviations.
As the plot shows, the mean value of the DFA exponent appears to be stable enough across different numbers of generated stimuli.
Also, the variance appears small enough for $S$ from $4,000$ on.
From such results, we decided to fix the $S = 10,000$, as it seems a good trade-off between experiments duration and DFA exponent stability.

\SubSection{SVM Training and Test}
In our experiments we employed a C-SVM with a radial basis function (RBF) kernel as the classifier. 
To build the training set we split our dataset of malware samples: 70\% for training, and remaining 30\% for the test.
For each training sample, we performed two independent executions, and thus we computed two fingerprints.
These fingerprints, labeled with the corresponding malware families, constitute the training set.
We repeated the same procedure with the test samples to build the test set.
We trained the SVM with $5$-fold cross-validation and evaluated our approach on the test set.
We repeated this procedure for 20 different 70\%-30\% random training/test partitions of the dataset.
We evaluated the accuracy of the classifier and finally compute the mean accuracy across these 20 repetitions.
During the training phase we follow the methodology reported in Section~\ref{sec:training} to determine the best subset of features to employ.

\SubSection{Results}

\begin{figure}[t]
	\centering
	\includegraphics[scale=0.3]{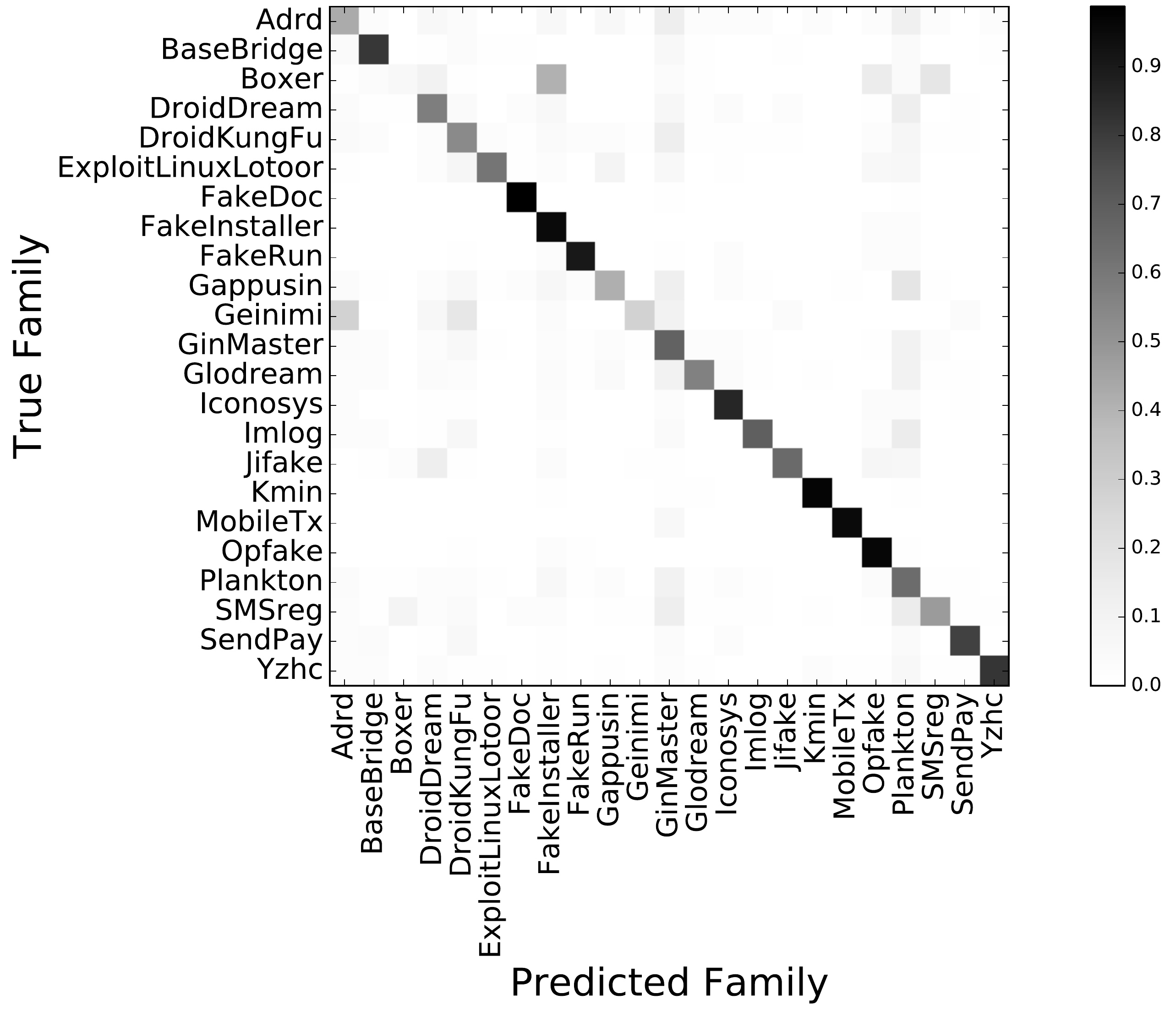}
	\caption{Normalized confusion matrix of the SVM classifier.}
	\label{figure:Confusion_matrix}
\end{figure}

We evaluate our methodology by assessing the accuracy of our classifier.
A commonly employed tool which gives a quick graphical overview of the performance of a classifier is the confusion matrix $M$.
The generic element $M_{ij}$ of the matrix is the number of samples belonging to class $i$ that has been classified as $j$ by the classifier.
Figure~\ref{figure:Confusion_matrix} shows the confusion matrix of our classifier.
Each element has been normalized in \mbox{0-1}, by dividing it by the sum of the related row, and the corresponding value as been codified with a gray scale for ease of visualization.

\begin{figure}[t]
\centering
\includegraphics[scale=0.18]{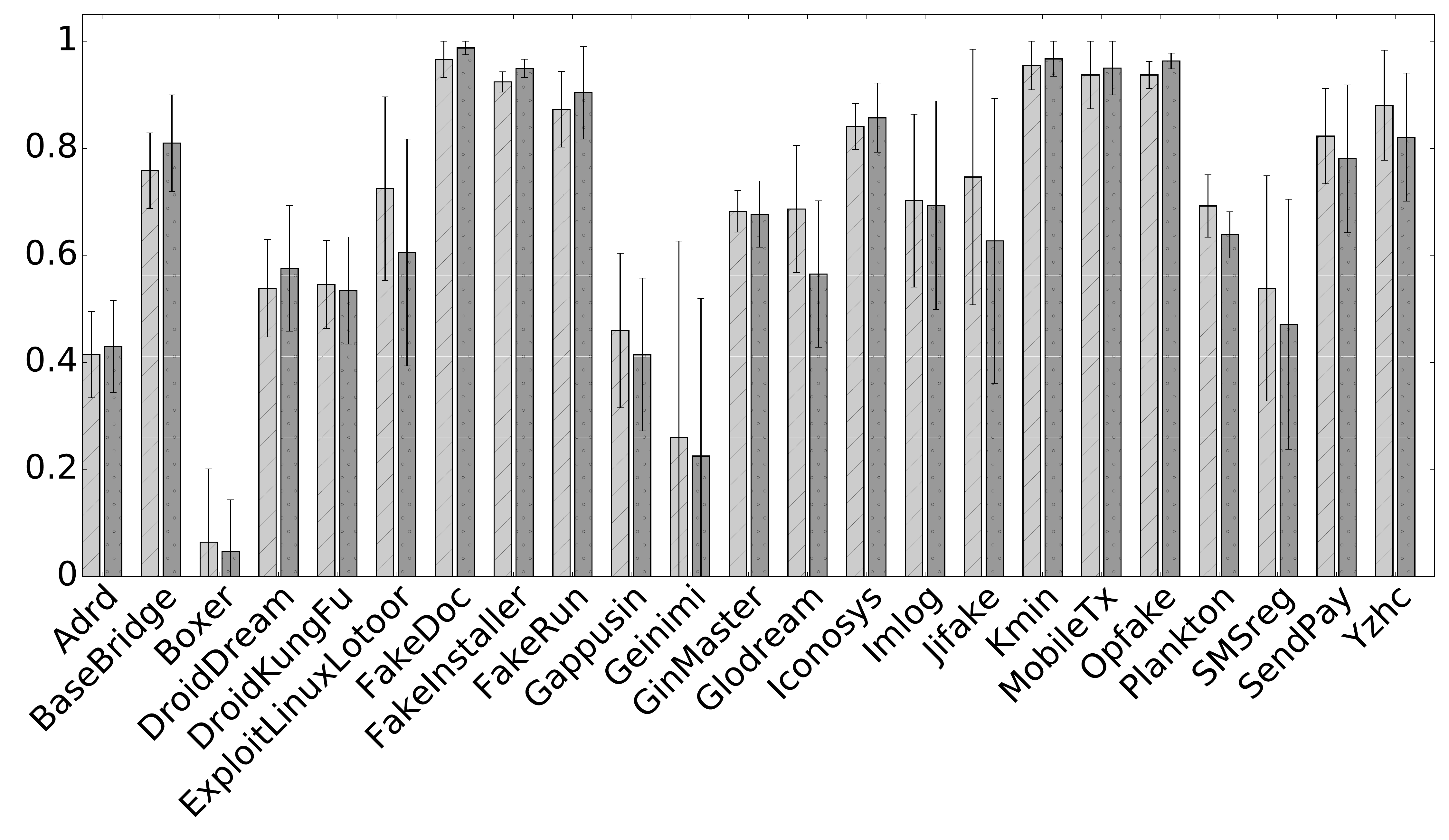}
\caption{Precision and recall values for all the 23 classes.}
\label{figure:precision_recall}
\end{figure}

To quantify numerically the performance of the classifier we employ three common metrics: accuracy, recall and precision.
%
%
The accuracy is an overall measure of the performance of the classifier, that can be defined from the confusion matrix $M$ as ${Accuracy = \frac{\sum_i M_{ii}}{\sum_i \sum_j M_{ij}}}$.
Precision and recall, instead, are class-related measures, and are defined as follows:
${Precision_i = \frac{M_{ii}}{\sum_j M_{ji}}}, {Recall_i = \frac{M_{ii}}{\sum_j M_{ij}}}$
%
After $20$ repetitions of our training and test methodology detailed in the previous section we obtained a mean value of $82\%$ for the accuracy with a standard deviation of $1\%$. 
In Figure~\ref{figure:precision_recall} we report mean and standard deviation values (depicted, respectively, as bars and whiskers) of precision and recall for each malware family.
From this figure and the confusion matrix depicted in Figure~\ref{figure:Confusion_matrix}, it is evident that some families achieves a nearly perfect classification (i.e. \textit{Fakedoc}, \textit{MobileTx}, \textit{Kmin}, \textit{Opfake}).
Conversely, two families, \textit{Boxer} and \textit{Geinimi}, are misclassified most of times.  
In particular, the first is often confused with the \textit{FakeInstaller} family. 
By analyzing some samples of this family we discovered that they contain the same activities as samples from the \textit{FakeInstaller} family. 
Moreover, by submitting these samples to VirusTotal\footnotemark, we noticed that some antiviruses classify them as \textit{Boxer.FakeInstaller}. 
This probably means that the behavior of samples of the \textit{Boxer} family is very similar to that of samples in the \textit{FakeInstaller} family.
Thus, our classifier is often misled. 
Instead, most of the samples belonging to the \textit{Geinimi} family crash at launch time or during the execution and we were able to collect data only for 7 applications of this family. 
This strong imbalance in the proportion of these samples in the training set could probably be the reason of the bad performance achieved for this class.

\SubSection{Comparison with DroidScribe}
We classify malware into families with a SVM as in DroidScribe.
Using a SVM classifier they are able to achieve 84\% accuracy on the Drebin dataset.
As already stated, we evaluate the accuracy for 20 different random splittings of the dataset into a training set and a test set, and then compute the mean value.
The authors of DroidScribe do not provide enough details about how the dataset was split.
Even though this prevent a precise comparison, our results (82\% accuracy) are similar achieved by DroidScribe.

\footnotetext{https://www.virustotal.com}DroidScribe also presents some results using conformal prediction to improve the accuracy of their approach.
In some cases, the confidence on the class output by classifiers (such as SVMs) can be very low. 
This may reduce the accuracy due to the fact that the classifier has to choose a single class even though the confidence is only slightly larger than the one on other classes.
Conformal prediction tries to mitigate this problem by letting the classifier output a prediction set, instead of a single class.
That is, a set of classes for which there is enough confidence that the set contains the true class.
In DroidScribe, they claim to be able to improve the accuracy from 84\% to 94\% by using conformal prediction.
They achieve 96\% accuracy on malware families for which the confidence is below a certain threshold.
However, in this case, the average prediction set size is $9$.
By including all malware families, the accuracy decreases to $94\%$ with a set size of $p>1$ ($p$~is not specified in their paper).
The value of the accuracy has to be necessarily related to the average prediction set size $\langle \textrm{accuracy}, \textrm{set-size} \rangle$ to be significant.
There is no evidence, in general, that $\langle 94\%, p \rangle$ is better than $\langle 84\%, 1 \rangle$. 
Thus, the two results are not directly comparable.
%
For these reasons, we compare to the solution of DroidScribe employing the SVM classifier without using conformal prediction. 

\Section{Conclusion}
\label{sec:conclusion}


In this work we have presented a novel methodology to classify Android malware into their belonging families.
The proposed approach is based on dynamic analysis and leverages resource consumption metrics collected during app execution.
Our approach is able to classify malware with an accuracy of 82\%.
This result is similiar to that obtained by DroidScribe, a state-of-the-art work which is the most related to ours.
Nevertheless, our methodology has the following advantages: (\textit{i}) it is easier to reproduce because only employs publicly available tools and does not require any modification to the emulated environment or Android OS, and (\textit{ii}) collected data can be gathered on physical devices.
In particular, the last point is crucial since many modern malware rely on evasion techniques to hide their malicious behavior on emulated environments.

As future work, we plan to increase the number of collected metrics that can effectively help in classifying malware families and test other machine learning algorithms to reach better accuracy results.
Moreover, we plan to evaluate our methodology on a more recent and larger dataset of malware samples, which has been recently released~\cite{Wei2017}.
Another aspect we intend to investigate is on-line learning.
This would allow to reduce the time required for updating the classifier and, thus, timely classify new malware families.
Finally, it would be interesting to test the effectiveness of our methodology on physical devices to be more robust against evasion techniques based on the detection of emulated environments. At this regard, we can investigate the possibility to integrate BareDroid~\cite{mutti2015baredroid} in the proposed architecture to reduce analysis overhead time.


\section*{Acknowledgments}
This present work has been partially supported by a grant of the Italian Presidency of Ministry Council, and by CINI Cybersecurity National Laboratory within the project \emph{FilieraSicura: Securing the Supply Chain of Domestic Critical Infrastructures from Cyber Attacks} (\url{www.filierasicura.it}) funded by CISCO Systems Inc. and Leonardo SpA.

\bibliographystyle{latex8}
\bibliography{references}

\end{document}